\documentclass[a4paper,11pt]{article}
\usepackage[margin=2cm]{geometry}
\usepackage{amssymb}
\usepackage{graphicx}
\usepackage{float}
\usepackage[section]{placeins}
\usepackage{tabularx,ragged2e,booktabs}
\usepackage{ctable}
\usepackage{multirow}
\usepackage{mathtools}
\usepackage[T1]{fontenc}
\usepackage[utf8]{inputenc}
\usepackage{hyperref}
\usepackage{doi}
\usepackage{cite}

\makeindex             

\begin{document}

\title{Regional Control of Probabilistic Cellular Automata}
\author{Franco Bagnoli$^{(1)}$, Sara Dridi$^{(2)}$, Samira El Yacoubi$^{(2)}$, Ra\'ul Rechtman$^{(3)}$\\[1cm]
   \noindent\begin{minipage}{0.7\columnwidth}
       \raggedright
1) Department of Physics and Astronomy and CSDC, \\University of Florence,\\ via G. Sansone 1, 50019 Sesto Fiorentino. Italy. \\Also INFN, sez. Firenze. \\\texttt{franco.bagnoli@unifi.it}\\
2) Team Project IMAGES\_ESPACE-Dev,\\  
UMR 228 Espace-Dev IRD UA UM UG UR,\\
University of Perpignan Via Domitia 
52, Avenue Paul Alduy. 
66860-Perpignan cedex. France.  
\\\texttt{yacoubi@univ-perp.fr }\\
3) Instituto de Energ\'\i{}as Renovables, \\Universidad Nacional Aut\'onoma de M\'exico, \\
Apartado Postal 34, 62580 Temixco, Morelos, Mexico. 
\\\texttt{rrs@ier.unam.mx}
\end{minipage}
}
\date{May 16, 2018}
\maketitle

\begin{abstract}
Probabilistic Cellular Automata are extended stochastic systems,  widely used for modelling phenomena in many disciplines. The possibility of controlling their behaviour is therefore an important topic. We shall present here an approach to the problem of controlling such  systems by acting only on the boundary of a target region.
\end{abstract}

\section{Introduction}
\label{sec:intro}
Cellular Automata (CA) are widely used for studying the mathematical properties of discrete systems and for modelling physical systems~\cite{Acri, Kauffman,Diversity,CAModelingBiologicalPatterns,CAApproachBioModeling,CACooperative}. They come in two major "flavours": deterministic CA (DCA)~\cite{Burks,Berlekamp,Vichniac,wolfram83,wolfram84,CASurvey} and probabilistic CA (PCA)~\cite{DK, pca}. 

DCA are the discrete equivalent of continuous dynamical systems (i.e., differential equations or maps) but are intrinsically extended, constituted by many elements, so they are in principle the discrete equivalent of system modelled by partial differential equations. 
DCA are defined by graph, a discrete set of states at the nodes of the graph, and a local transition function that gives the future state of a node as a function of the present state of the node connected to it, its so-called neighbourhood.  This evolution rule is applied in parallel to all nodes. 
PCA can be thought as an extension of DCA where the transition function gives the probability that the target node goes in a certain state. If all these probabilities are either zero or one, that the PCA reduces to a DCA. In both cases, the state of the CA
is the collection of states at the nodes of the graph
and this state changes in time according to functions defined in every node of the graph.

In analogy with continuous dynamical systems, it is important to develop methods for controlling the  behaviour of DCA and PCA. In particular, the main control problems for extended systems are reachability and drivability. The first is related to the possibility of applying a suitable control able to make the system reach a given state or a set of states. For instance, assuming that the system under investigation represents a population of pests, the control problem could be that of bringing the population towards extinction at a given time or to keep the population under a certain threshold. 

The drivability problem is somehow complementary to the reachability one; once that the system is driven to a desired state or collection of states, what kind of control may make it follow a given trajectory? For instance, one may want to stabilize a fixed point, or make the system follow a cycle, and so on.

As usual in control problems, one aims at achieving the desired goal with the optimal cost or smallest effort, and we speak of an optimal control problem. One may be  interested not in controlling the whole space, but rather the state of a given region, for instance how to avoid that a pollutant reaches a  certain area. 

The techniques for controlling discrete systems are quite different from those used in continuous ones, since discrete systems are in general strongly non-linear and the usual linear approximations cannot be directly applied. What one can do is to change the state at a node or a set of chosen nodes. For Boolean CA the 
state is either 0 or 1, so a change is either 1 or 0. The ``intensity'' of the control therefore can be only associated to the average number of changes, and cannot be made arbitrary small. We are interested in regional control of PCA, that is, how to achieve a certain goal in a set of neighbouring nodes of a graph.

This problem is related to the so-called regional controllability introduced in Ref.~\cite{Zerrik}, as a special case of output controllability~\cite{Lions,Lions1,Russell}. The regional control problem consists in achieving an objective only in a subregion of the domain when some specific  actions are exerted on the system, in its domain interior or on its boundaries.  This  concept has been studied by means of partial differential equations. Some  results on the action properties (number, location, space distribution) based on the rank condition  have been obtained depending on the target region and its geometry,  see for example Ref.~\cite{Zerrik} and the references therein. 

Regional controllability has also been studied using CA models. In Ref.~\cite{Samira}, a numerical approach based on genetic algorithms has been developed for a class of additive CA in in one and two dimensions. In Ref.~\cite{regionalca}, an interesting theoretical study has been carried out for one dimensional additive CA where the effect of control is given through an evolving neighbourhood and a very sophisticated state transition function.
However, these studies did not provide a real insight in the regional controllability problem.

Some results for control techniques applied to one dimensional DCA can be found in Refs.~\cite{Samira1, bagnoli10, controlPRE, Bagnoli-natcomp, Bagnoli-Ency}.

For DCA, once the states in the neighbouring nodes are known, the future state at the node under consideration is fixed and for PCA we have in general only the probability of reaching a certain state. One advantage of PCA vs. DCA is that their dynamics can be fine-tuned. PCA are summarized in Sec.~\ref{sec:PCA}.

The control problem of PCA is more subtle than of DCA. In general, it is impossible to exactly drive these systems towards a given configuration, but it is possible to increase the probability that the system 
will reach a target state in a collection of nodes, or, alternatively, to lower as much as possible the probability of the appearance of a given configuration, for instance the extinction of a species inside a given region. 

The  evolution of a PCA can be seen as a Markov chain, where the elements of the transition matrix are given by the product of the local transition probabilities (Sec.~\ref{sec:PCA}). In particular we shall study here a particular PCA (BBR model) with two absorbing states in Sec.~\ref{sec:BBR}.

A Markov chain is said to be ergodic if there is the possibility of going to any state in the graph to any other state in a finite number of steps. If this goal can be achieved for all pairs of states at a given time, the Markov chain is said to be regular. This consideration allows us to define the reachability problem in terms of the probability, once summed over all possible realizations of the control, of connecting any two sites. And since DCA can be considered as the extreme limit of PCA, this technique can be applied to them too, see Sec.~\ref{sec:Reachability}. 

Finally, one should remark that the problem of controllability (in particular that of drivability) is strictly related to that of synchronization (see Ref.~\cite{controlPRE} for instance). In this same issue the regional synchronization problem for the BBR model is addressed~\cite{BagnoliRechtman-RegionalSynchronization}.

\section{Definitions}\label{sec:definitions}

Cellular Automata are defined on graph composed by $N$ nodes identified by an index $i=1, \dots, N$,
by an adjacency matrix $a_{ij}$ that establishes the neighbourhood of each node with $a_{ij} =1$
($a_{ij} =0$) if node $j$ is (is not) in node $i$'s neighbourhood, and by a transition function $f_i$
that gives the new state at node $i$ given the states in its neighbourhood.
The connectivity of node $i$ is $k_i =\sum_j a_{ij}$. We shall deal here with graphs having fixed connectivity $k_i=k$ and use the same transition function in all the nodes, $f_i=f$.

A lattice is a graph invariant by translation and the nodes are called sites. For a one dimensional lattice  with $N$ sites with connectivity $k=2r+1$, $r=1,2,\dots$ and $r$ the range, the neighbourhood
of site $i$ is the set $\{i-r,\dots,i+r\}$. We impose
Periodic boundary conditions are generally imposed.
The state at site $i$ at time $t$, $x_i(t)$, is chosen from a finite set of values, for Boolean CA,
$x_i(t)\in \{0,1\}$. Then
\[
 x_i(t+1)=f(x_{i-r}(t),\dots,x_{i+r}(t))
\]
 
On each node $i$ of the graph there is one dynamical variable $x_i=x_i(t)$ that for Boolean CA  only takes values 0 and 1. 
We shall indicate with $x'_i=x_i(t+1)$ its value at the following time step. 

\begin{figure}[t]
\begin{center}
\begin{tabular}{cc}
\begin{minipage}{0.6\columnwidth}
\includegraphics[width=\columnwidth]{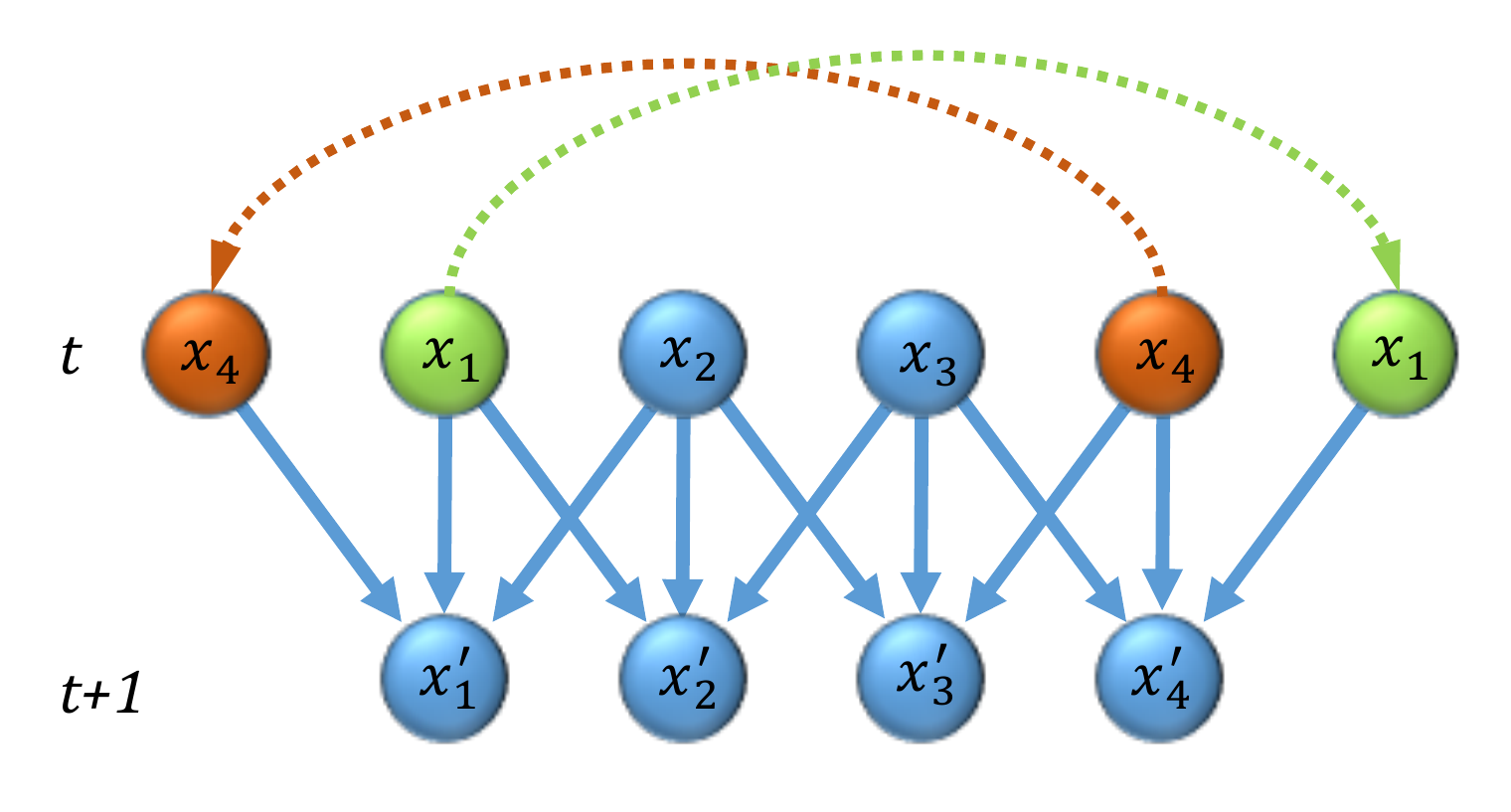}
\end{minipage}
&
\begin{minipage}{0.4\columnwidth}
\includegraphics[width=\columnwidth]{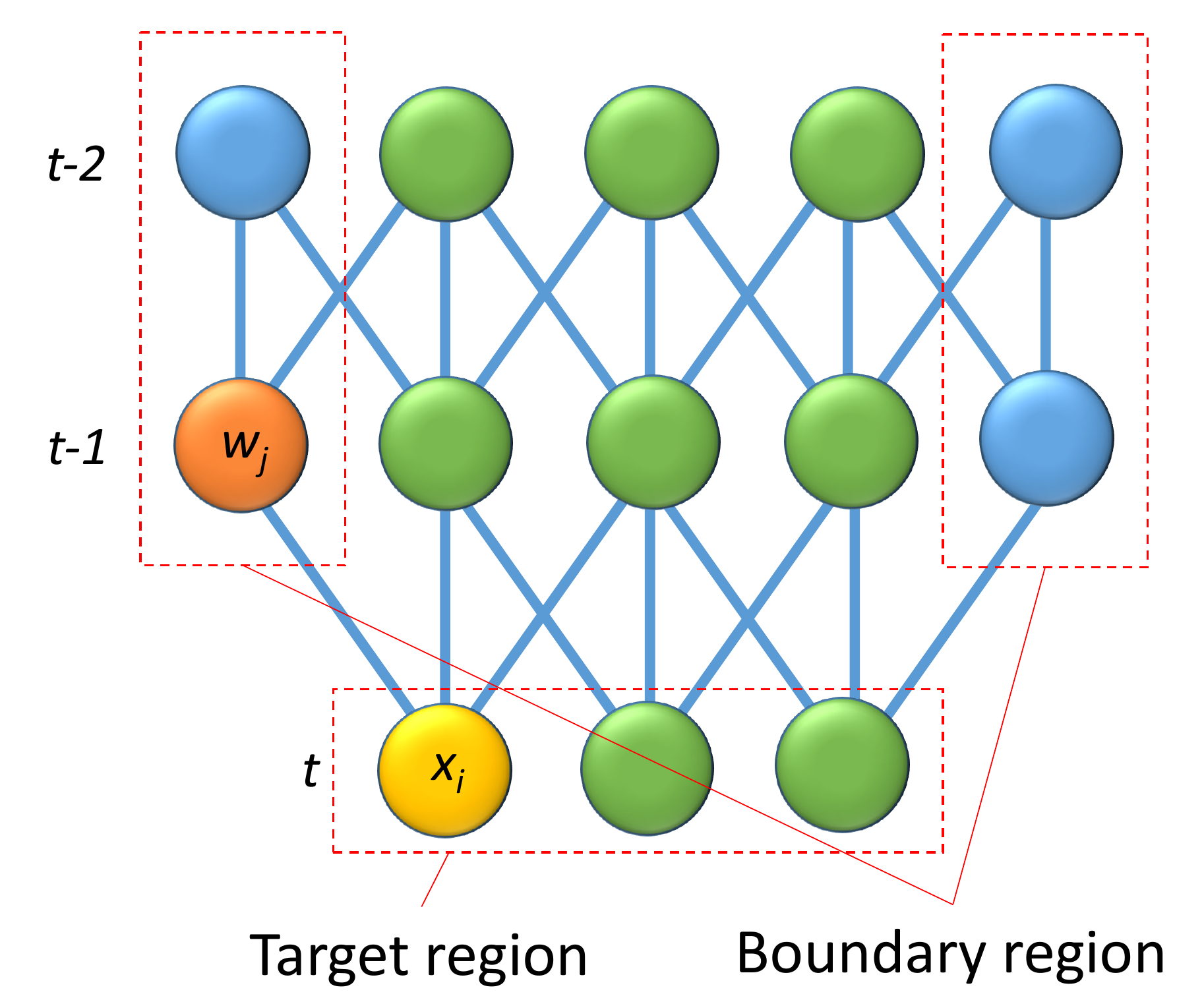}
\end{minipage}
\end{tabular}
\end{center}
\caption{\label{fig:lattice} Left: The space-time lattice of 1D CA with periodic boundary conditions. Right: CA  boundary-value  problem.}
\end{figure}

An ordered set of Boolean values like  $x_1, x_2, \dots, x_N$ can be read as a Boolean vector or as base-two number and we shall indicate it as $\boldsymbol{x}$, $0\le \boldsymbol{x}<2^N$. 
We shall also indicate with $\boldsymbol{v}_i$ the  state of all connected neighbours. The state of $x'_i$ depends on the state of the neighbourhood $\boldsymbol{v}_i$, and  on some random number $r_i(t)$ for stochastic CA. In formulas (neglecting to indicate the random numbers) we have 
\[
x'_i = f(\boldsymbol{v}_i).
\]
The function $f$ is applied in parallel to all sites. Therefore, we can define a vector function $\boldsymbol{F}$ such that 
\[
    \boldsymbol{x}' = \boldsymbol{F}(\boldsymbol{x}).
\]
The sequence of states $\{\boldsymbol{x}(t)\}_{t=0,\dots}$ is a trajectory of the system with $\boldsymbol{x}(0)$ as the initial condition. 

When  $f$ depends symmetrically on the states of neighbours, it can be shown that $f$ actually depends on the sum $s_i = \sum_j a_{ij} x_j$.   
In this case we say that the cellular automaton is totalistic and write 
\begin{equation}
x_i(t+1)=f_T(s_i(t)),
\end{equation}
with $f_T:\{0,\dots,k\} \to \{0,1\}$. Totalistic cellular automata are generic, since they exhibit the whole variety of behaviour of general rules~\cite{wolfram83}. 
It is possible to visualize the evolution of the automata as happening on a space-time oriented graph or lattice, Fig.~\ref{fig:lattice}-left.

\section{Probabilistic Cellular Automata}\label{sec:PCA}
Probabilistic CA constitute an extension of DCA. Let us introduce the transition probability $\tau(1|\boldsymbol{v})$ that, given a certain configuration $\boldsymbol{v}=\boldsymbol{v}_i$ of the neighbourhood of site $i$, gives the probability of observing $x'_i=1$ at next time step. Clearly $\tau(0|\boldsymbol{v})=1-\tau(1|\boldsymbol{v})$. DCA are such that $\tau(1|\boldsymbol{v})$ is either 0 or 1, while for PCA it can take any value in the middle. For a PCA with $k$ inputs, there are $2^k$ independent transition probabilities, and for totalistic PCA there are $k+1$ independent probabilities. If one associate each transition probability to a different axis, the space of all possible PCA is an unit hypercube, with corners corresponding to DCA.

\begin{figure}[t]
    \begin{center}
        \begin{tabular}{cc}
            \includegraphics[width=0.45\columnwidth]{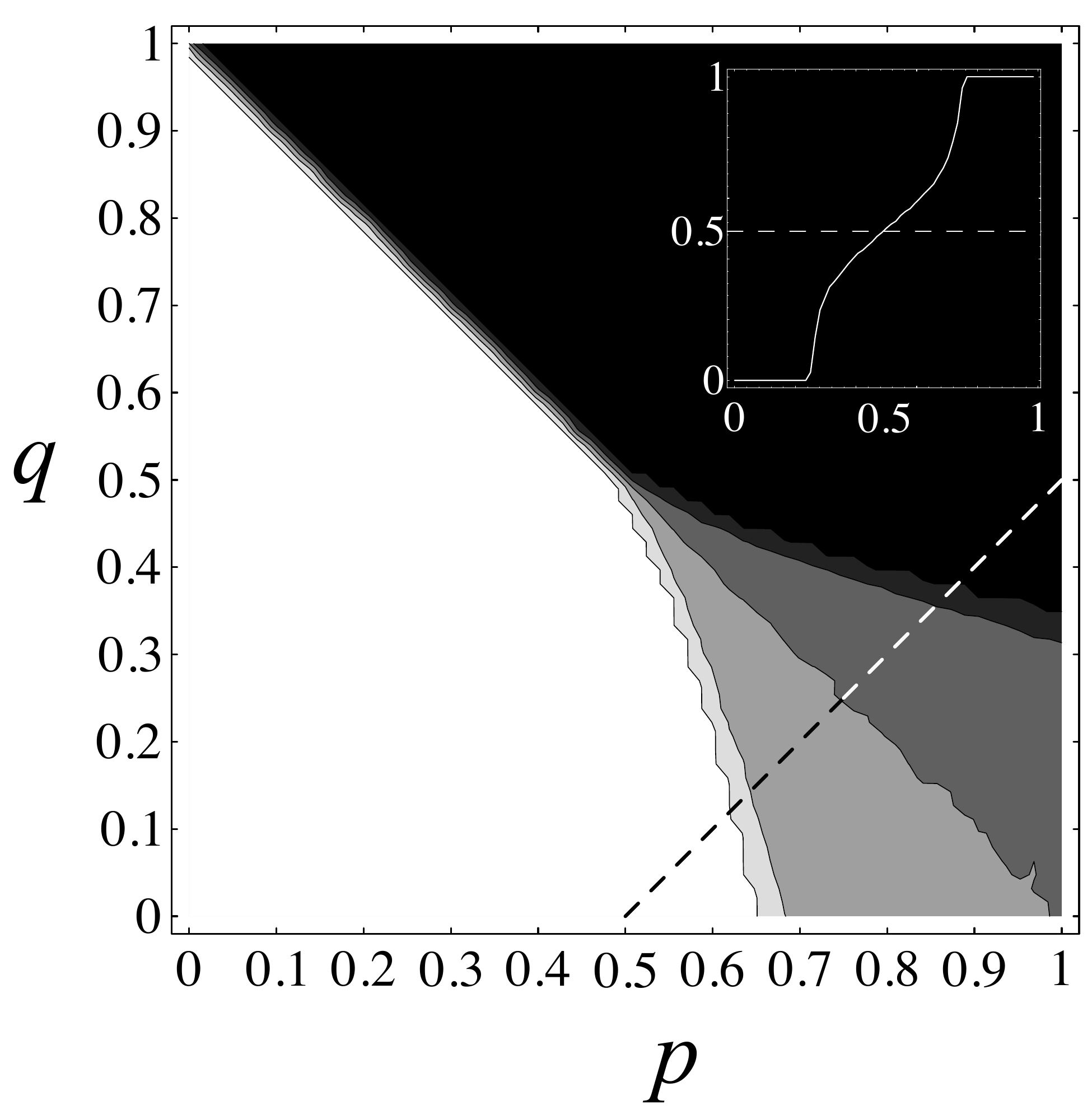}&
            \includegraphics[width=0.44\columnwidth]{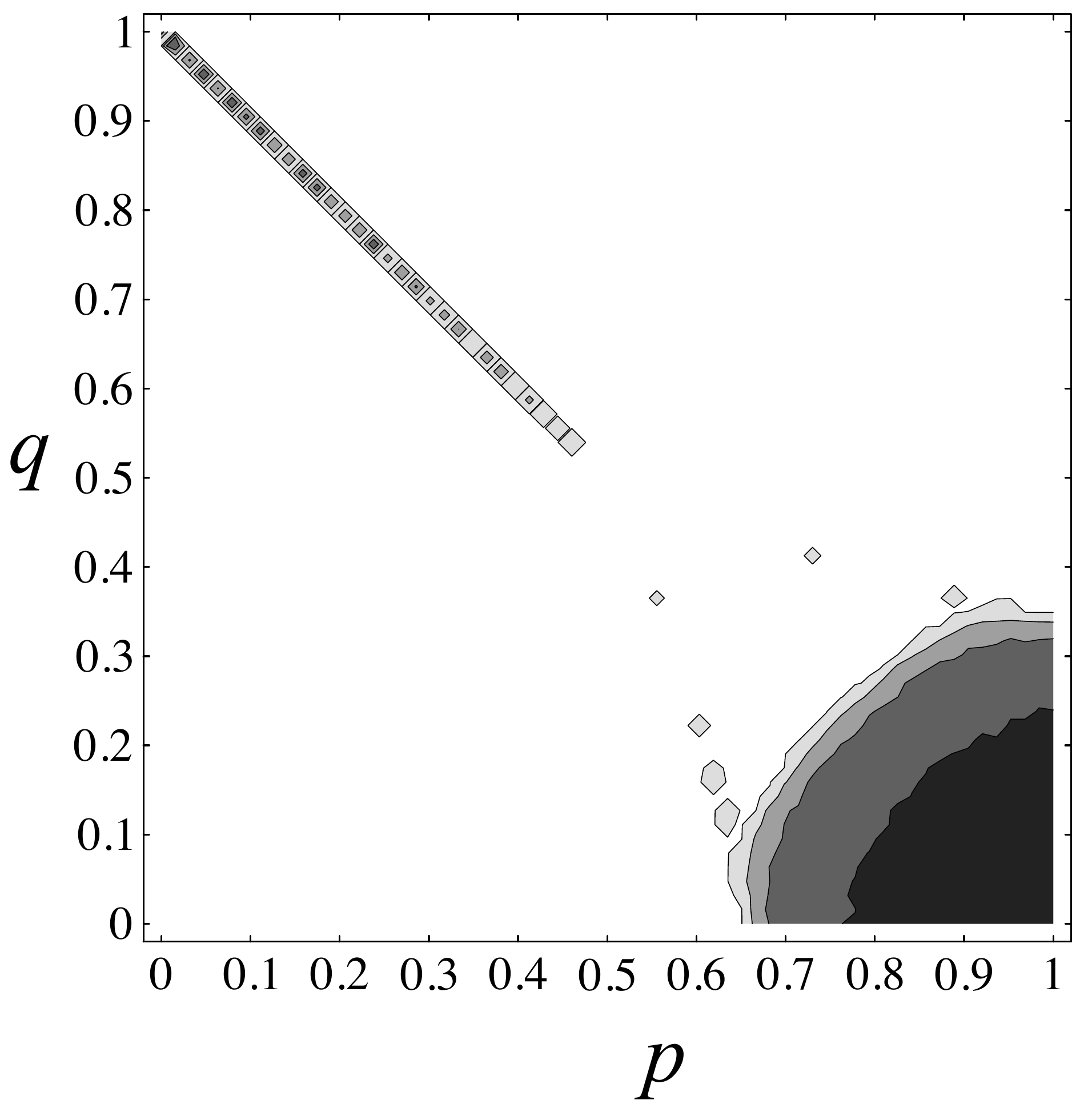}
        \end{tabular}
        \caption{\label{fig:3inp} Phase diagram of the BBR model. Left: Density phase diagram. Right: Damage phase diagram.}
    \end{center}
\end{figure}

PCA can be also \textit{partially} deterministic, i.e., the transition probability $\tau(1|\boldsymbol{v})$ can be zero or one for certain $\boldsymbol{v}$. This opens the possibility for the automata to have one or more absorbing state, i.e., configurations that always originate the same configuration (or give origin to a cyclic behaviour). The BBR model illustrated below has one or two absorbing states.

The evolution of all possible configurations $\boldsymbol{x}$ of a PCA can be written as a Markov chain. Let us define the probability $P(\boldsymbol{x}, t)$, i.e., the probability of observing the configuration $\boldsymbol{x}$ at time $t$. Its evolution is given by
\begin{equation}
P(\boldsymbol{x}, t+1) = \sum_{\boldsymbol{y}} M(\boldsymbol{x}|\boldsymbol{y}) P(\boldsymbol{y},t),
\end{equation}
where the matrix $M$ is such that 
\begin{equation}
M(\boldsymbol{x}|\boldsymbol{y}) = \prod_{i=1}^N \tau\left(x_i|\boldsymbol{v}_i(\boldsymbol{y})\right).
\end{equation}

For a CA on a 1D lattice and $k=3$ we have 
\begin{equation}
M(\boldsymbol{x}|\boldsymbol{y}) = \prod_{i=1}^N \tau(x_i|y_{i-1}, y_i, y_{i1}).
\end{equation}

Phase transitions for PCA can be described as degeneration of eigenvalues in the limit $N\rightarrow\infty$ and (subsequently) $T\rightarrow \infty$~\cite{Bagnoli-CA}.

Notice that since DCA are limit cases of PCA, they also can be seen as particular Markov chains.

A Markov chain such that, for some $t$, $(M^t)_{ij}>0$ for all $i,j$ is said to be regular, and this implies that any configuration can be reached by any configuration in time $t$. A weaker condition (ergodicity) says that $t$ may depend on the pair $i, j$ (for instance, one may have an oscillating behaviour such that certain pairs can be connected only for even or odd values of $t$). Also for ergodic systems all configurations are connected. 

\begin{figure}[t]
    \begin{center}
        \begin{tabular}{cc}
            \includegraphics[width=0.45\columnwidth]{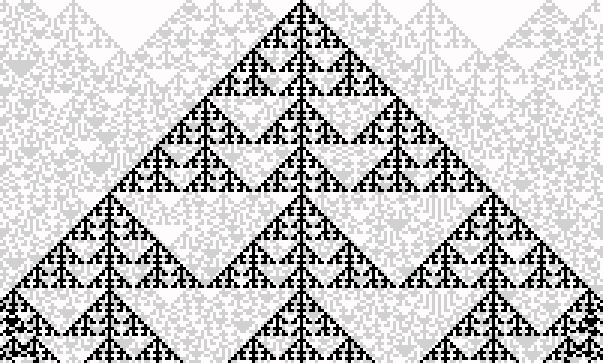}&
            \includegraphics[width=0.45\columnwidth]{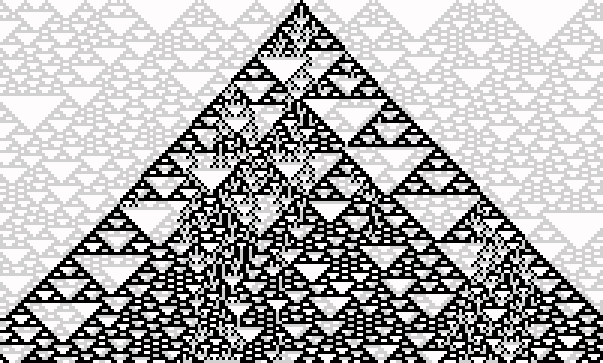}
        \end{tabular}
        \caption{\label{fig:damage} Damage spreading; time runs downwards. Left: CA Rule 150. Right: CA rule 126.}
    \end{center}
\end{figure}

\section{The model}\label{sec:BBR}
We shall use as a testbed model the one presented in Ref.~\cite{BBR}, which is an extension of the Domany-Kinzel CA~\cite{DK}. We shall refer to it as the BBR model from the name of its authors. It is a totalistic PCA defined on a one-dimensional lattice, with connectivity $k=3$. The transition probabilities of the model are
\begin{equation}
 \tau(1|0)=0; \qquad \tau(1|1)=p; \qquad \tau(1|2)=q; \qquad \tau(1|3)=w.
\end{equation}  
This model has one absorbing state, corresponding to configuration $\boldsymbol{0}=(0,0,0,\dots)$, For $w=1$ also the configuration $\boldsymbol{1} = (1,1,1,\dots)$ is an absorbing state. This is the version studied in Ref.~\cite{BBR}. 
 
Notice that for $p=1$, $q=1$, $w=0$ we have DCA rule  126 while for $p=1$, $q=0$ , $w=1$ we have DCA rule 150. In the following we shall use $w=1$.

The implementation of a stochastic model makes use of one of more random numbers. For instance, the BBR model can be implemented using the function
\begin{equation}
\begin{split}
x'_i = f(x_{i-1}, x_i, x_{i+1}; &r_i) = [r_i < p](x_{i-1}\oplus x_i\oplus x_{i+1} \oplus x_{i-1}x_i  x_{i+1} ) \\
&\oplus [r_i < q](x_{i-1}x_i\oplus x_{i-1}x_{i+1}\oplus x_ix_{i+1} \oplus x_{i-1}x_i  x_{i+1})   \\
&\oplus x_{i-1}x_i  x_{i+1},
\end{split}
\end{equation}
where  $[\cdot]$ is the truth function which takes value one if $\cdot$ is true and zero otherwise, and $\oplus$ is the sum modulo two. The $r_i = r_i(t)$ random numbers have to be extracted for each site and for each time. One can think of extracting them once and for all at the beginning of the simulation, i.e., running the simulation on a space-time lattice on which a  random field $r_i(t)$, $i=1,\dots,N$; $t=0,\dots$ is defined. Notice that in this way one has a deterministic CA over a quenched random field.  

The phase diagram of the BBR model is reported in Fig.~\ref{fig:3inp}-left. One can see three regions. The one marked in white, for $p<0.65$, is where the only asymptotically stable configuration is the absorbing state formed by all zeros, i.e., the asymptotic probability distribution of configurations $P(\boldsymbol{x})$ is a delta on zero. The symmetric region marked in black, for $q>0.35$ is where the only stable configuration is formed by all ones. Actually, in a region near the diagonal $q=1-p$, for $p<0.5$ the two absorbing states are both stable, the transition line is fixed by the initial configuration, which in the figure is drawn at random with the same probability of extracting a zero and a one. These regions are denoted with the term ``quiescent''. The region marked in shades of grey, for $p>0.65$ and $q<0.35$ is a region where the two absorbing states are unstable, and the asymptotic probability distribution is distributed over many configurations, with average number of ones proportional to the shades of grey. In the insect it is reported the asymptotic average  number of ones (the ``density'') computed along the dashed lines. This region is denoted with the term ``active''.

\subsection{Damage spreading} \label{sec:damage}

One possibility for controlling the evolution of a system with little efforts is offered by the sensitive dependence on initial conditions, i.e., when a small variation in the initial state propagates to the whole system. Indeed, this is also the main ingredient of chaos, which in general prevents a careful control. But in discrete systems the situation is somehow different. These systems are not affected by infinitesimal perturbations in the variables (assuming that they can be extended in the continuous sense), only to finite ones. The study of the propagation of a finite perturbation in CA goes under the name of ``damage spreading'', indicating how an initial disturbance (a ``defect'' or ``damage'') can spread in the system. A CA where a damage typically spreads is said to be chaotic. 

Mathematically, one has two copies of the same system, say $x$ and $y$, evolving with the same rule but starting from different initial conditions. We shall indicate with $z_i = x_i \oplus y_i$ the local difference at site $i$. Typical patterns of the spreading of a damage (i.e., the evolution of $z$) are reported in Fig.~\ref{fig:damage}

For PCA, the concept of damage spreading is meant ``given the random field''. The phase diagram of the damage $z$ for the BBR model is shown in Fig~\ref{fig:3inp}-right.

\section{Reachability problem}\label{sec:Reachability}

We shall mainly deal here with the problem of regional control via boundary actions, i.e., boundary reachability as illustrated in Fig.\ref{fig:lattice}-right, however the techniques of analysis can be extended to other cases.

Let us now consider the problem of computing the probability $M_{\boldsymbol{x}\boldsymbol{y}}(a,b) = M(\boldsymbol{x}|\boldsymbol{y}; a,b)$ which is the probability of getting configuration $x$ at time $t+1$ given the configuration $y$ at time $t$, and boundaries $a$ and $b$ (for simplicity we refer here only to one-dimensional cases). The Markov matrix $M(a,b)$ is given by 
\[
M_{\boldsymbol{x}\boldsymbol{y}}(a,b) = \tau(x_1|a,y_1,y_2)\tau(x_2|y_1,y_2, y_3)\dots\tau(x_n|y_{n-1},y_n, b),
\]
where $n$ indicates the size of the target region. 

For a given control sequence $\boldsymbol{a}=a_1, \dots, a_T$ and $\boldsymbol{b}=b_1, \dots, b_T$, the resulting Markov matrix for time $T$ is 
\[
M(\boldsymbol{a}, \boldsymbol{b}) = \prod_{t=1}^T M(a_t, b_t).
\]

We can define several control problems. A first one is about ergodicity: which is the best control sequence $\boldsymbol{a}$ and $b$ so that $M_{\boldsymbol{x}\boldsymbol{y}}(\boldsymbol{a}, \boldsymbol{b})>0$ for all pairs $\boldsymbol{x}, \boldsymbol{y}$ and minimum time $T$? Another is: given a certain time $T$ and a pair   $\boldsymbol{x}, \boldsymbol{y}$, which is the best control sequence $\boldsymbol{a}$ and $\boldsymbol{b}$ that maximises $M_{\boldsymbol{x}\boldsymbol{y}}(\boldsymbol{a}, \boldsymbol{b})>0$? 

Clearly, one can also be interested in avoiding certain configurations, for instance, if $x_i=1$ represents the presence of some animal or plant in position $i$ at time $t$, one could be interested in devising a control that prevents the extinction of animals, i.e., avoid the state $\boldsymbol{x}=0$. 

As we shall show in the following, so far we have not found algorithms for finding the best control but exhaustive search. 

Beyond finding the actual sequence that maximises the observable, one could be rather  interested in determining  the \textit{existence} of such a sequence, for a certain time interval $T$, or to find the minimum time $T$ for which an optima sequence exists.

In particular this latter problem can be faced with less computer efforts than finding the actual sequence for the best control. If one considers the matrix
\[
C= \frac{1}{4} \sum_{a,b} M(a,b) = \frac{1}{4} \bigl(M(0,0)+M(0,1)+M(1,0)+M(1,1)\bigr),
\]
and then computes its power $C^T$, all possible control sequences of length $T$ are contained in such a power. Therefore, the problem of the existence of a control sequence for a given time $T$  reduces to checking if $(W^T)_{\boldsymbol{x}\boldsymbol{y}} > 0$. One can also quantify the effective of the  control by computing the ratio $\eta$ between the minimum and maximum values of $C$. If this ratio is zero, it means that there are certain pairs of configurations that cannot be connected by any control sequence, while  $\eta=1$ means that all pairs of configurations can be connected with equal easiness. 

\begin{figure}[t]
    \begin{center}
        \begin{tabular}{cc}
        \includegraphics[width=0.5\columnwidth]{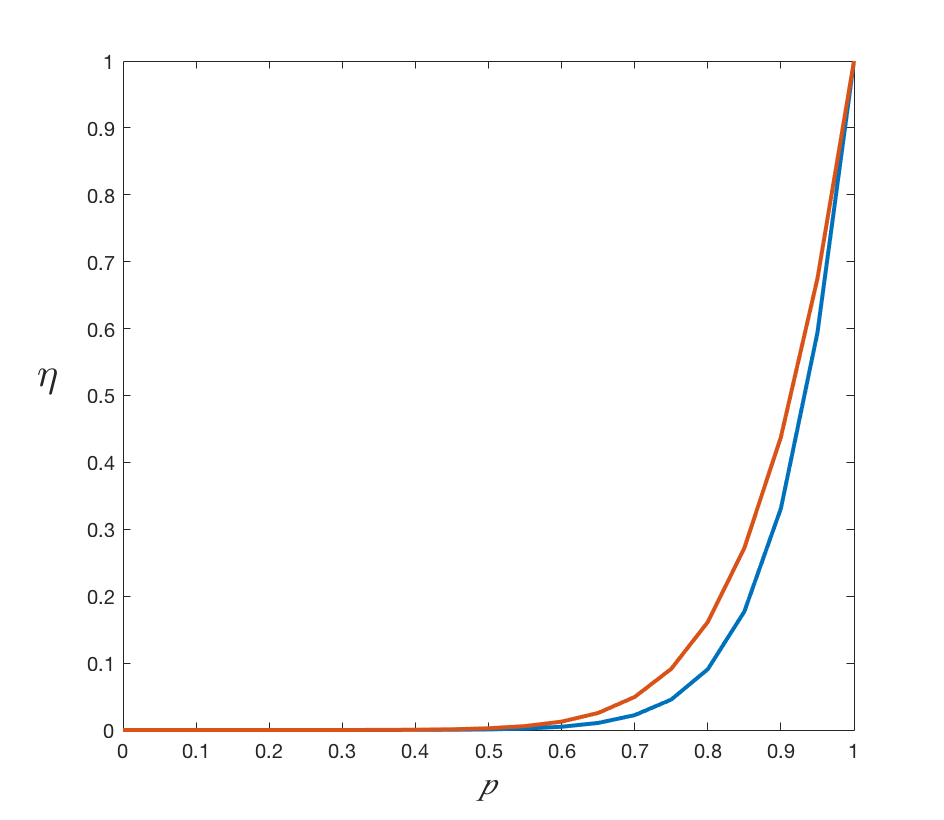}&
        \includegraphics[width=0.5\columnwidth]{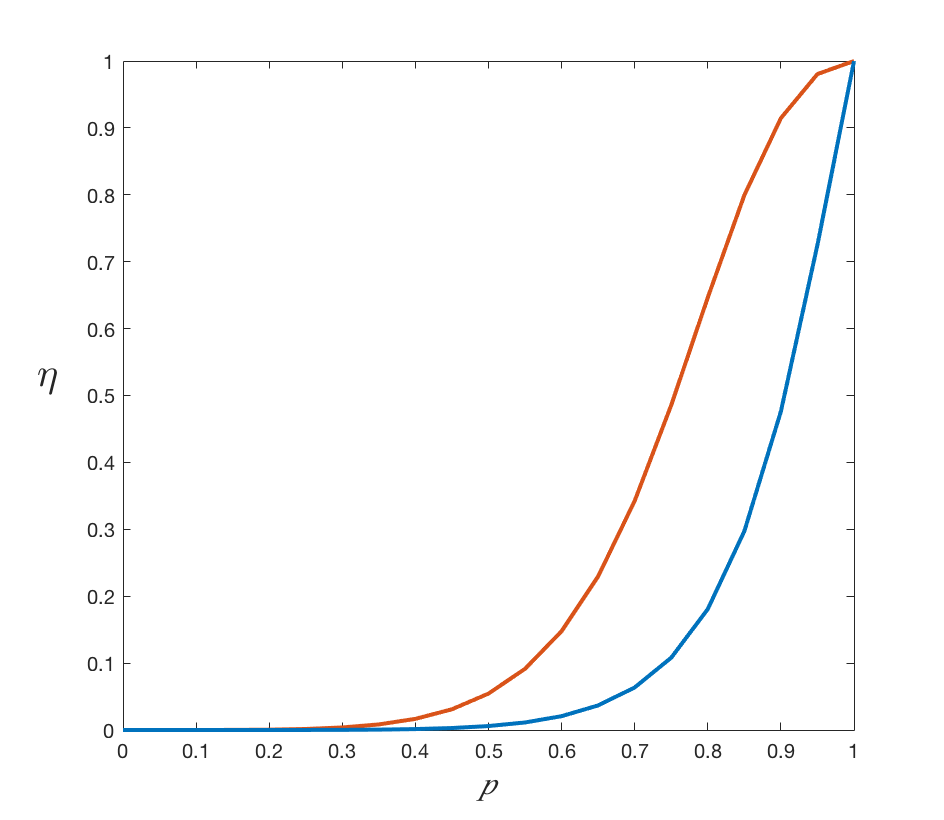}
    \end{tabular}
        \caption{\label{fig:ControlBBR} The ratio $\eta=\min(C)/\max(C)$ for the BBR model with $n=5$ for $T=3$ (lower, blue curve) and $T=5$ (upper, red curve). Left: $q=0$, Right: $q=1-p$}
    \end{center}
\end{figure}

Let us illustrate some of these concepts for the BBR model, for $p=q$ and for $q=0$. In Fig.~\ref{fig:ControlBBR} we show the easiness  parameter $\eta$ in function of $p$ for $q=0$ and $q=1-p$, for $n=5$ and different values of  $T$. One can see that in the ``quiescent'' phase $p<0.5$ the control is almost impossible, and that on the line $q=1-p$, for $p>0.5$, the easiness of the control rises with $T$ faster that on the line $q=0$. Indeed, referring to Fig.~\ref{fig:3inp}, one can see that this portion of the diagram corresponds to the ``active'' phase, where the BBR model is ergodic. One can also notice that the easiness of the control is not related to the damage spreading phase: considering for instance the line $q=p$, from Fig.~\ref{fig:3inp}-right one sees that the damage spreading phase starts for $p>0.75$, while from Fig.~\ref{fig:ControlBBR}-right one sees that the control is possible well before this threshold. The control properties are probably associated to the ``chaoticity'' of the associated deterministic CA over the random quenched field, a problem which will be faced in the future (for ``chaotic'' CA and the associated Boolean derivatives, see Refs.~\cite{vichniac84,bagnoli92,bagnoli99}).

Let us now turn to the problem of finding the best control.  For compactness, let us consider the case $n=3$, for which the minimum control time  is $T=2$. The highest probability for each pair of configurations $\boldsymbol{x}$ (row index in base two) and $\boldsymbol{y}$ (column index in base two) for $q=1-p$ and $p=0.7$ is
\[ M=
\bordermatrix{
    ~ &  $0$ & $1$ & $2$ & $3$ & $4$ & $5$ & $6$ & $7$ \cr
    $0$ & $1.000$ & $0.262$ & $0.213$ & $0.396$ & $0.262$ & $0.396$ & $0.396$ & $0.240$ \cr
    $1$ & $0.700$ & $0.278$ & $0.208$ & $0.293$ & $0.208$ & $0.293$ & $0.293$ & $0.343$ \cr
    $2$ & $0.343$ & $0.221$ & $0.221$ & $0.253$ & $0.221$ & $0.195$ & $0.253$ & $0.490$ \cr
    $3$ & $0.343$ & $0.293$ & $0.293$ & $0.278$ & $0.293$ & $0.208$ & $0.208$ & $0.700$ \cr
    $4$ & $0.700$ & $0.208$ & $0.208$ & $0.293$ & $0.278$ & $0.293$ & $0.293$ & $0.343$ \cr
    $5$ & $0.490$ & $0.253$ & $0.195$ & $0.221$ & $0.253$ & $0.221$ & $0.221$ & $0.343$ \cr
    $6$ & $0.343$ & $0.293$ & $0.293$ & $0.208$ & $0.293$ & $0.208$ & $0.278$ & $0.700$ \cr
    $7$ & $0.240$ & $0.396$ & $0.396$ & $0.262$ & $0.396$ & $0.213$ & $0.262$ & $1.000$
},
\]
corresponding to controls $\boldsymbol{a}$ and $\boldsymbol{b}$ (again in base two) 
\[ a=
\bordermatrix{
    ~ &  $0$ & $1$ & $2$ & $3$ & $4$ & $5$ & $6$ & $7$ \cr
    $0$ & $0$ & $1$ & $1$ & $0$ & $0$ & $1$ & $1$ & $1$ \cr
    $1$ & $2$ & $2$ & $2$ & $2$ & $3$ & $3$ & $3$ & $3$ \cr
    $2$ & $0$ & $0$ & $0$ & $3$ & $1$ & $1$ & $1$ & $1$ \cr
    $3$ & $1$ & $2$ & $2$ & $1$ & $3$ & $0$ & $0$ & $3$ \cr
    $4$ & $0$ & $3$ & $3$ & $0$ & $2$ & $1$ & $1$ & $2$ \cr
    $5$ & $2$ & $2$ & $1$ & $2$ & $0$ & $3$ & $3$ & $0$ \cr
    $6$ & $0$ & $0$ & $0$ & $0$ & $1$ & $1$ & $1$ & $1$ \cr
    $7$ & $1$ & $2$ & $2$ & $3$ & $3$ & $2$ & $2$ & $3$
}
\qquad \boldsymbol{b}=
\bordermatrix{
    ~ &  $0$ & $1$ & $2$ & $3$ & $4$ & $5$ & $6$ & $7$ \cr
    $0$ & $0$ & $0$ & $1$ & $1$ & $1$ & $1$ & $0$ & $2$ \cr
    $1$ & $0$ & $2$ & $3$ & $1$ & $3$ & $1$ & $0$ & $2$ \cr
    $2$ & $3$ & $1$ & $0$ & $1$ & $0$ & $2$ & $3$ & $1$ \cr
    $3$ & $0$ & $1$ & $0$ & $1$ & $0$ & $1$ & $0$ & $1$ \cr
    $4$ & $2$ & $3$ & $2$ & $3$ & $2$ & $3$ & $2$ & $3$ \cr
    $5$ & $2$ & $0$ & $2$ & $3$ & $2$ & $3$ & $2$ & $3$ \cr
    $6$ & $1$ & $3$ & $2$ & $0$ & $2$ & $0$ & $1$ & $3$ \cr
    $7$ & $2$ & $3$ & $2$ & $2$ & $2$ & $2$ & $3$ & $3$
}.
\]

These results should be read in this way. Let us consider for instance the initial configuration $y=3=110|_2$ (numbers are coded in reverse order) and final configuration $x=4=001|_2$. The best control is given by a sequence $a=0=00|_2$ and $b=3=11|_2$, which is reasonable since one is trying to force zeros on the left side of the configurations and ones on the right side.

Notice however that the entries for $\boldsymbol{a}$ and $\boldsymbol{b}$ are not always either 0 or 3, meaning that the best control is not a uniform one for all pairs. For instance, for going from $y=3=110|_2$ to $x=1=100|_2$  one has to apply  $a=2=01|_2$ and $b=1=10|_2$, exploiting the fact that $q=\tau(1|3) =1-p=0.3$ and therefore for forcing a zero in the presence of a neighbourhood already containing a one, it is better to insert another one than a zero.

\section{Conclusions and future perspectives}\label{sec:Conclusions}
We have introduced the problem of  controlling probabilistic cellular automata by an action performed on the boundary of a target region (boundary control or boundary reachability problem). We have formulated the problem and presented the first results. 

The field of control of cellular automata and discrete systems is extremely recent and only  a handful of results are known~\cite{Bagnoli-natcomp,Bagnoli-Ency}. In particular, the control of probabilistic cellular automata is still to be explored in depth, and more efficient  algorithms  for finding the best control sequence are needed if one wants to exert control on large regions, and in higher dimensions. 

A promising possibility is that of exploring the relationship between the control and the ``chaotic'' properties of the associated deterministic CA over a quenched random field.

\end{document}